\documentclass[twocolumn,showpacs,amsmath,amssymb,aps,prb,floatfix,nofootinbib,superscriptaddress]{revtex4}
\usepackage{graphicx,graphics,times}
\usepackage{bm}

\begin{document}

\title{Level Crossing Analysis of the Stock Markets}

\author{G. R. Jafari}
\affiliation{Department of Physics, Shahid Beheshti University,
Evin, Tehran 19839, Iran} \affiliation{Computational Physical
Sciences Research
Laboratory Department of Nano-Science,\\
Institute for studies in theoretical Physics and Mathematics (IPM),
P. O. Box 19395-5531, Tehran, Iran}

\author{M. S. Movahed}
 \affiliation{Department of Physics, Sharif University of Technology,
P.O. Box 11365-9161, Tehran, Iran} \affiliation{Institute for
studies in theoretical Physics and Mathematics (IPM), P. O. Box
19395-5531, Tehran, Iran}

\author{S. M. Fazeli}
\affiliation{Department of Physics, Sharif University of Technology,
P.O. Box 11365-9161, Tehran, Iran}

\author{M. Reza Rahimi Tabar}
 \affiliation{Department of Physics, Sharif University of Technology,
P.O. Box 11365-9161, Tehran, Iran}
 \affiliation{ CNRS UMR 6529, Observatoire de
la C$\hat o$te d'Azur, BP 4229, 06304 Nice Cedex 4, France}

\author{S. F. Masoudi}
\affiliation{Department of Physics, K.N Toosi University of
Technology, P.O. Box 15875-4416, Tehran, Iran}

\begin{abstract}
We investigate the average frequency of positive slope
$\nu_{\alpha}^{+}$, crossing for the returns of market prices.
 The method is based on stochastic processes which no scaling
feature is explicitly required. Using this method we define new
quantity to quantify stage of development and activity of stocks
exchange. We compare the Tehran and western stock markets and show
that some stocks such as Tehran (TEPIX) and New Zealand (NZX) stocks
exchange are emerge, and also TEPIX is a non-active market and
financially motivated to absorb capital.
\end{abstract}
\maketitle

\section{Introduction}

In recent years, financial markets have been at focus of
physicists's attempts to apply existing knowledge from statistical
mechanics to economic problems
\cite{Friedrich,Sornette,Ausloos,Stanley,Rice,Wang,Sarma,Kim,Johansen,Sergio}.
Statistical properties of price fluctuations are very important to
understand and model financial market dynamics, which has long been
a focus of economic research. These markets, though largely varying
in details of trading rules and traded goods, are characterized by
some generic features of their financial time series. The aim is to
characterize the statistical properties of given time series with
the hope that a better understanding of the underlying stochastic
dynamics could provide useful information to create new models able
to reproduce experimental facts. An important aspect concerns the
ability to define concepts of activity and degree of development of
the markets. Acting on advantageous information moves the price such
that the a priori gain is decreased or even destroyed by the
feedback of the action on the price. This makes concrete the concept
that prices are made random by the intelligent and informed actions
of investors, as put forward by Bachelier, Samuelson, and many
others \cite{Sornette}. In contrast, without informed traders, the
profit opportunity remains, since the buying price is unchanged.
Based on recent research for characterizing the stage of development
of markets \cite{Payam,Matteo1,Matteo2}, it is well known that the
Hurst exponent shows remarkable differences between developed and
emerging markets.

Here we introduce "level crossing" to analyze these time series.
Which is based on stochastic processes grasps the scale dependency
of time series
\cite{doe87,Tabar1,Mogens1,Mogens2,Mogens3,Mogens4,Mogens5} and no
scaling feature is explicitly required. Also this approach has
turned out to be a promising tool for other systems with scale
dependent complexity (see \cite{Tabar1,Tabar2} for it is application
to characterize the roughness of growing surfaces). Some authors
have applied this method to study the fluctuations of velocity
fields in Burgers turbulence \cite{Movahed}, and in the
Kardar-Parisi-Zhang equation in d + 1-dimensions
\cite{Bahraminasab}.

The level crossing analysis is very sensitive to correlation when
the time series is shuffled and to probability density functions
(PDF) with fat tails when the time series is surrogated. The long
range correlations are destroyed by the shuffling procedure and in
the surrogate method the phase of the discrete Fourier transform
coefficients of time series are replaced with a set of
pseudo-independent distributed uniform $(-\pi,+\pi)$ quantities. The
correlations in the surrogate series do not change, but the
probability function changes to Gaussian distribution
\cite{sc00,the92,the93,the97}.

 Level crossing with detecting correlation is a useful tool
to find the stage of development of markets, too. It is known that
emerging markets have long-range correlation. This sensitivity of
level crossing to the market conditions provides a new and simple
way of empirically characterizing the development of financial
markets. This means that mature markets the total level crossing are
decreased under shuffling effectively, while emerging markets are
increased in agreement with the findings of Di Matteo et al (2003)
and (2005) which indicate that emerging markets have H = 0.5, while
mature markets have H = 0.5 [H is the Hurst exponent]. The level
crossing analysis is more simple calculation than the other methods
such as, Detrended Fluctuation Analysis (DFA)
\cite{Peng94,murad,physa,kunhu,kunhu1,dns,sad06}, Detrended Moving
Average (DMA) \cite{ales}, Wavelet Transform Modulus Maxima (WTMM)
\cite{wtmm}, Rescaled Range analysis (R/S) \cite{hurst65,eke02},
Scaled Windowed Variance (SWV) \cite{eke02} etc.  It is well known
that, R/S, SWV and other non-detrending methods work well if the
records are long and do not involve trends. Also in the detrending
method one must make attention that, in some cases, there exist one
or more crossover (time) scales separating regimes with different
scaling exponents \cite{physa,kunhu,sad06}. In this case
investigation of the scaling behavior is more complicate and
different scaling exponents are required for different parts of the
series \cite{kunhu1}. Therefore one needs a multitude of scaling
exponents (multifractality) for a full description of the scaling
behavior. Crossover usually can arise either because of changes in
the correlation properties of the signal at different time (space)
scales, or can often arise from trends in the data.  The level
crossing analysis does not require  modulus maxima procedure in
contrast with WTMM method, and hence does not require lot's of
effort in to write computing code  and computing time, than the
above methods. So the level crossing analysis is more suitable for
short time series.

This paper is organized as follows. In section II we discuss the
level crossing in detail. Data description and analysis based on
this method for some stocks indices are given in section III.
Section IV closes with a discussion of the present results.

\section{Level Crossing Analysis}

Let us consider a time series $\{p(t)\}$, of price index with length
$n$ , and the price returns $r(t)$ which is defined by $r(t) = \ln
p(t+1)-\ln p(t)$. Here, we investigate  the detrended $\log$ returns
for different time scales.

 Let for time interval T $\nu_{\alpha}^{+}$ denotes the number of positive
difference crossing $r(t)- \bar r = \alpha$  (see fig.$1$) and also
mean value for all the samples be $N_{\alpha}^{+}(L)$ where
\begin{equation}
N_{\alpha}^{+}(T)=E[n_{\alpha}^{+}(T)].
\end{equation}
Since after detrending $r(t)$, is  stationary (i.e. averaged
variance saturate to a certain value) if we take  second consecutive
time interval  $T$  we  obtain the same result, and therefore for
two intervals together we have
\begin{equation}
N_{\alpha}^{+}(2T)=2N_{\alpha}^{+}(T),
\end{equation}
from which it follows that, for  stationary process, the average
number of crossing is proportional to the space interval $T$. Hence
\begin{equation}
N_{\alpha}^{+}(T)\propto T,
\end{equation}
or
\begin{equation}
N_{\alpha}^{+}(T)=\nu^{+}_{\alpha} T.
\end{equation}
which $\nu_{\alpha}^{+}$ is the average frequency of positive
slope crossing of the level $r(t)- \bar r = \alpha$.
  We show how the frequency parameter
$\nu_{\alpha}^{+}$ can be deduced from the underlying probability
distribution function PDF for $r(t) - \bar r$. In time interval
$\triangle t$
 the sample can only cross $r(t) - \bar r=\alpha$ with positive
difference if it has the property $r(t)-\bar r < \alpha$ at the
beginning of this time interval. Furthermore there is a minimum
difference at time $t$ if the level $r(t)- \bar r = \alpha$ is to be
crossed in interval $\triangle t$ depending on the value of $r(t)-
\bar r$ at time $t$. So there will be a positive crossing of
$r(t)-\bar r =\alpha$ in the next time interval $\triangle t$ if, at
time $t$,
\begin{figure}[t]
\includegraphics[width=8cm,height=7cm,angle=0]{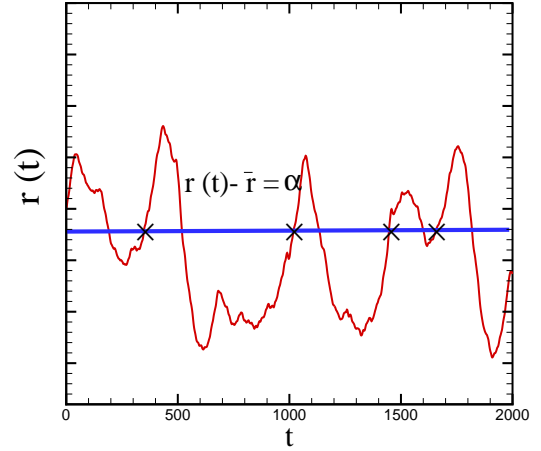}
\caption{positive slope crossing in a fixed $\alpha$ level.}
\end{figure}
\begin{equation}
r(t)- \bar r < \alpha \hspace{.6cm} and \hspace {.6cm}
\frac{\triangle\left[r(t)-\bar r\right]}{\triangle t}>
\frac{\alpha-\left[r(t) - \bar r\right] }{\triangle t}.
\end{equation}
As shown in \cite{Tabar1} the frequency $\nu_{\alpha}^{+}$ can be
written in terms of joint PDF of $p(\alpha,{y}^{\prime})$ as follows
\begin{equation}\label{level}
\nu_{\alpha}^{+}=\int_{0}^{\infty}p(\alpha,{y}^{\prime}){y}^{\prime}d{y}
^{\prime}.
\end{equation}
where ${y}^{\prime} = \frac{ r(t+\Delta t)- r(t)}{\Delta t}$. Here
we put $\Delta t =1$.
\begin{figure}[t]
\includegraphics[width=8cm,height=7cm,angle=0]{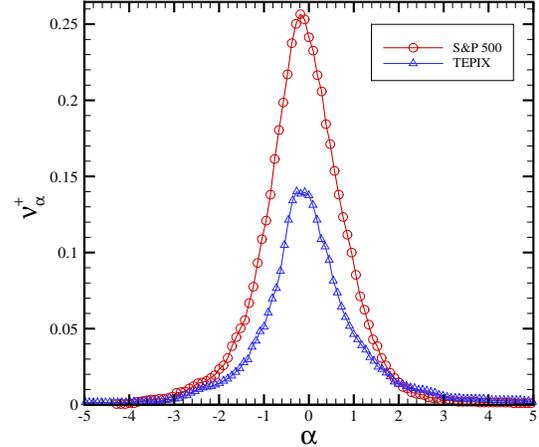}
\caption{The positive difference crossing of return price for
 S\&P500 and TEPIX market in the same time
 interval.}\label{fig2}
\end{figure}

Let us also introduce the quantity $N_{tot}^{+}(q)$ as
 \begin{equation}\label{ntq}
N_{tot}^{+}(q)=\int_{-\infty}^{+\infty}\nu_{\alpha}^{+} |\alpha -
\bar{\alpha}|^{q} d \alpha.
\end{equation}
where zero moment (with respect to $\nu_{\alpha}^{+}$) $q=0$, shows
the total number of crossing for return price with positive slope.
The moments $q<1$ will give information about the frequent events
while moments $q>1$ are sensitive for the tail of events.

\begin{figure}[t]
\includegraphics[width=8cm,height=7.5cm,angle=0]{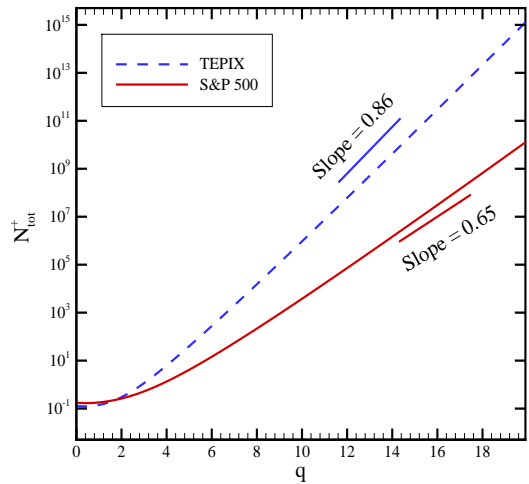}
\caption{Generalized total number of crossing with positive slope
$N_{tot}^{+}$ for TEPIX and S$\&$P $500$ market.}\label{fig3}
\end{figure}
\begin{figure}
\includegraphics[width=8cm,height=13cm,angle=0]{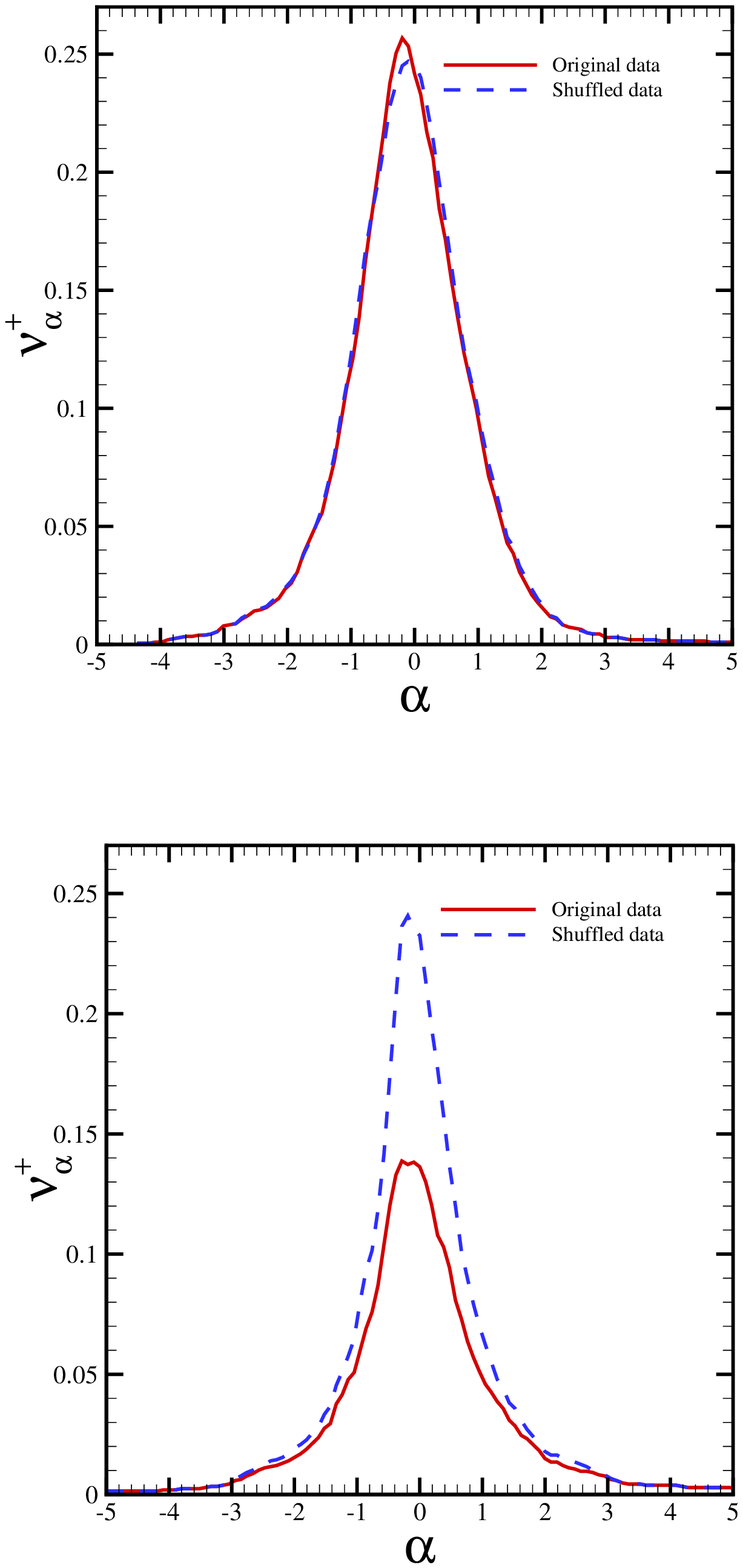}
\caption{Comparison the positive slope crossing of return price
between
 original and shuffled data for S$\&$P $500$ (upper panel) and TEPIX
 (lower panel) market in the same time interval.}\label{fig4}
\end{figure}

\section{Application on Stock Market}

Investments in the stock market are based on a quite straightforward
rule: if you expect the market to go up in the future, you should
buy (this is referred to as being "long" in the market) and hold the
stock until you expect the trend to change direction; if you expect
the market to go down, you should stay out of it, sell if you can
(this is referred to as being "short" of the market) by borrowing a
stock and giving it back later by buying it at a cheaper in the
future. It is difficult, to say the least, to predict future
directions of stock market prices even if we are considering time
scales of the order of decades, for which one could hope for a
negligible influence of "noise."

The reason why, in very liquid markets of equities and currency
exchanges correlations of returns are extremely small, is because
any significant correlation would lead to an arbitrage opportunity
that is rapidly exploited and thus washed out. Indeed, the fact that
there are almost no correlations between price variations in liquid
markets can be understood from simple calculation by
\cite{Sornette,ro}. In other words, liquidity and efficiency of
markets control the degree of correlation, that is compatible with a
near absence of arbitrage opportunity. It is important to consider,
that the more intelligent and hard working the investors, the more
random is the sequence of price changes generated by such a market.

Acting on advantageous information moves the price such that the a
priori gain is decreased or even destroyed by the feedback of the
action on the price. This makes concrete the concept that prices are
made random by the intelligent and informed actions of investors, as
put forward by Bachelier, Samuelson, and many others
\cite{Sornette}. In contrast, without informed traders, the profit
opportunity remains, since the buying price is unchanged. Grossman
and Stiglitz \cite{gross} argued that perfectly informationally,
efficient markets are an impossibility, for if markets are perfectly
efficient, the return on gathering information is nil, in which case
there would be little reason to trade and markets would eventually
collapse. Alternatively, the degree of market inefficiency
determines the effort investors are willing to expend to gather and
trade on information, hence a non degenerate market equilibrium will
arise only when there are sufficient profit opportunities, that is,
inefficiencies, to compensate investors for the costs of trading and
information-gathering. The profits earned by these industrious
investors may be viewed as economic rents that accrue to those
willing to engage in such activities. Who are the providers of these
rents? noise traders, individuals who trade on what they think is
information but is in fact merely noise. More generally, at any time
there are always investors who trade for reasons other than
information (for example, those with unexpected liquidity needs),
and these investors are willing to "pay up" for the privilege of
executing their trades immediately.

For these purposes, we have analyzed the  level crossing of
detrended log return time series, $r(t)$ for
 S$\&$P$500$, Djindu, Biojen, $10$ytsy, Composite, Amex, TEPIX and NZX. To have a good comparison,
 we have chosen the time series from the same time
interval: 20 May 1994 to 18 March 2004, and data have been recorded
at each trading day. Consider a price time series with length $n$.
Here, we investigate on the detrended log returns on different time
scales. To remove the trends present in the time scales ${r(t)}$ in
each subinterval of length s, we fit $r(t)$ using a linear function,
which represents the exponential trend of the original index in the
corresponding time window. After this detrending procedure, we
define detrended log returns, $r(t)$ is a deviation from the fitting
function \cite{Kiyono1, Kiyono2}. According to Eq. \ref{level} the
level crossing, $\nu^{+}_{\alpha}$, is calculated for each index.
Figure \ref{fig2} shows a comparison of $\nu^{+}_{\alpha}$ for TEPIX
and S$\&$P$500$ as a function of level $\alpha$. It is clear that
$\nu^{+}_{\alpha}$ scales inversely with time, so
$\tau(\alpha)=\frac{1}{\nu^{+}_{\alpha}}$ is a time interval, within
this time, the level crossing in average will be observed again.
Table \ref{Tb1} shows the time interval in the high frequency
($\tau(\alpha=0)$) and the low frequency (tails,
$\tau(\alpha=\pm3\sigma)$) regimes for some indices. The time
interval $\tau(\alpha=0)$ of TEPIX and S$\&$P$500$ are $7.0$ and
$4.0$ days, respectively. Still, in the tails, it is comparable.
Another difference between TEPIX and the other markets (except Amex
market) is seen in the time interval of left
($\tau(\alpha=-3\sigma)$) and right ($\tau(\alpha=+3\sigma)$) tails.
The time length in left tail is larger than time length in right but
also less than other markets and also in TEPIX and NZX the mean
$\overline{\alpha}$ is $0.24$ and $0.17$, respectively. They mean
that TEPIX is financially motivated to absorb capital. It is clear
that when we apply Eq. \ref{ntq} for small $q$ regime, high
frequency is more significant, whereas in the large $q$ regime, low
frequency (the tail) is more significant. Figure \ref{fig3} shows
that when $q < 2$, the value of $N^+_{tot}$ for TEPIX is smaller
than that of S$\&$P$500$, while for $q>2$, the value of $N^+_{tot}$
for TEPIX gets larger than the other markets. This is because for
small $q$ the low frequency events of tails are more significant
than the high frequency peak.
 According to the last section and Eq. \ref{ntq}, the area under the level crossing
curve, $N^{+}_{tot}(q=0)$, shows the total number of crossing. This
means that the larger the area, the larger the activity. In essence,
by comparing $N^{+}_{tot}(q=0)$, activity of the index is obtained.

From another point of view, based on recent research of
characterizing stage of development of markets
\cite{Matteo1,Matteo2,Payam}, it is shown that the Hurts exponent is
sensitive to the degree of development of the market. Emerging
markets are associated with high value of Hurts exponent and
developed markets are associated with low value of the exponent. In
particular, it is found that all emerging markets have Hurts
exponents larger than $0.5$ (strongly correlated) whereas all the
developed markets have Hurts exponents near to or less than 0.5
(white noise or anti-correlated). Here we have shown that the level
crossing has ability to characterize degree of development of
markets. The sensitivity of the level crossing to the market
conditions provides a new and more simple way of empirically
characterizing activity and development of financial markets.

Since $N^{+}_{tot}(q=0)$ is very sensitive to correlation, it
increases when the time series is shuffled so that the correlation
disappears. Thus, by comparing the change between $N^{+}_{tot}(q=0)$
and $N^{+}_{sh}(q=0)$ (shuffled), the stage of development of
markets can be determined. Figure \ref{fig4} shows $\nu^+_{\alpha}$
as a function of $\alpha$ for original and shuffled data in TEPIX
and S$\&$P$500$. The relative changing of $N^{+}_{tot}(q=0)$ for
TEPIX and S$\&$P$500$ are $0.41$ and $0.02$ respectively. For the
sake of comparison between various stock markets, the two parameters
$N^{+}_{tot}$ and $N^{+}_{sh}$, relative variation of
$N^{+}_{tot}(q=0)$ for original data, its shuffled and Hurst
exponent which was obtained by using the detrended fluctation
analysis (DFA) method \cite{Peng94}, are reported in Table
\ref{Tb2}. We notice that TEPIX and NZX belong to the emerging
markets category; it is far from an efficient and developed market.
These result are comparable with the results reported in
\cite{Payam} and show that Tehran stock exchange belongs to the
category of emerging financial markets. The level crossing analysis
is more simple calculation than the other methods such as
generalized Hurst exponent approach, DFA , rescaled range analysis
(R/S), wavelet techniques (WT) etc. Also in short time series these
methods are not stable etc.

\begin{table}
\caption{\label{Tb1} The values of $\tau(\alpha)$ for different
level, $\alpha$.}
\medskip
\begin{tabular}{|c|c|c|c|}
  \hline
  Market & $\tau(\alpha=0)$ & $\tau(\alpha=3\sigma)$ & $\tau(\alpha=-3\sigma)$ \\\hline
  S$\&$P 500  & 4.0 & 218.4& 115.1 \\\hline
  Djindu & 4.0 &  178.3 & 150.0  \\\hline
  Biogen &4.0  & 179.1  & 113.8 \\\hline
  10ytsy & 4.2& 656.3& 98.2  \\\hline
  Composit &  4.3& 178.6& 140.1  \\\hline
  Amex &4.6& 115.1& 178.4  \\\hline
   NZX &5.3& 135.3& 120.3  \\\hline
  TEPIX & 7.0& 102.8& 114.2  \\\hline
\end{tabular}
\end{table}

\begin{table}
\caption{\label{Tb2}The values of $N^{+}_{tot}(q=0)$,
$N_{sh}^{+}(q=0)$ and Hurst exponent for some markets during the
same period.}
\medskip
\begin{tabular}{|c|c|c|c|c|}
  \hline
  Market & $N_{tot}^{+}$ & $N_{sh}^{+}$ & $|N_{sh}^{+}-N_{tot}^{+}|/N_{tot}^{+}$& $H$ \\\hline
  S$\&$P 500  & 0.52 & 0.53& 0.02& 0.44$\pm$ 0.01  \\\hline
  Djindu & 0.51 &  0.52 & 0.02& 0.46 $\pm$ 0.01  \\\hline
  10ytsy & 0.50&  0.52& 0.04& 0.47 $\pm$ 0.01  \\\hline
  Biogen &0.48  & 0.51  & 0.06& 0.51 $\pm$ 0.01 \\\hline
  Composit &  0.50& 0.52& 0.04& 0.45$\pm$ 0.01  \\\hline
  Amex &0.45& 0.50& 0.10&0.51$\pm$ 0.01 \\\hline
  NZX &0.40& 0.52& 0.30&0.61$\pm$ 0.01 \\\hline
  TEPIX & 0.32& 0.45& 0.41& 0.74$\pm$ 0.01  \\\hline
\end{tabular}
\end{table}

\section{Conclusion}

In this paper concept of level crossing analysis has been applied to
several stock market indices. It is shown that the level crossing is
able to detect activity of markets. This method is based on
stochastic processes which should grasp the scale dependency of any
time series in a most general way. No scaling feature is explicitly
required. Based on the recent research for characterizing the stage
of development of markets \cite{Matteo1,Matteo2,Payam}, it is shown
that level crossing is sensitive to degree of development of market,
too. This sensitivity of level crossing to market conditions
provides a new and simple way of empirically characterizing the
activity and development of financial markets. Considering all of
the above discussions and results, we notice that Tehran Stock
Exchange belongs to the emerging markets category. It is far from an
efficient and developed market and also we have found that it is
financially motivated to absorb capital. Using this method we
classify the activity and stage of development of some markets.

\section{Acknowledgment}

Jafari would like to acknowledge the hospitality extended during his
visits at the IPAM, UCLA, where this work was done. We would like to
thank D. Sornette for his useful comments and K. Ghafoori-Tabrizi
and T. J. Stasevich for reading the manuscript.


\end{document}